\documentclass[twocolumn,twoside,slac_two]{revtex4}
\usepackage{graphicx}
\usepackage{fancyhdr}
\begin{document}

\title{Coronal Electron Scattering of Hot Spot Emission Around Black Holes}

%

\author{Jeremy D. Schnittman}
\affiliation{Massachusetts Institute of Technology, Cambridge, MA 02139, USA}

\begin{abstract}
Using a Monte Carlo ray-tracing code in full general relativity, we
calculate the transport of photons from a geodesic hot spot emitter
through a corona of hot electrons surrounding a black hole. Each
photon is followed until it is either captured by the black hole or is
detected by a distant observer. The source is assumed to be a
low-energy thermal emitter ($T_{\rm em} \sim 1$ keV), isotropic in the
rest frame of
a massive geodesic test particle. The coronal scattering has two major
observable effects: the Comptonization of the photon spectrum due to
the high energy electrons, and the convolution of the time-dependent
light curve as each photon is effectively scattered into a different
time bin. Both of these effects are clearly present in the {\it Rossi X-Ray
Timing Explorer (RXTE)} observations of high frequency quasi-periodic
oscillations (QPOs) seen in black hole binaries. These QPOs tend to
occur when the system is in the Steep Power Law spectral state and also show no
evidence for significant power at higher harmonic frequencies,
consistent with the smoothing out of the light curve by multiple
random time delays. We present simulated photon spectra and light
curves and compare with {\it RXTE} data, allowing us to infer the
properties of the corona as well as the hot spot emitter.

\end{abstract}

\maketitle

\thispagestyle{fancy}


\section{INTRODUCTION}
In recent years, observations of accreting black holes with the {\it
Rossi X-ray Timing Explorer (RXTE)} have discovered a number of sources with
high frequency quasi-periodic oscillations (HFQPOs) in their X-ray
light curves. These sources are seen predominately in the Steep
Power Law (SPL) spectral state, suggesting a significant level of
inverse-Compton scattering of the emitted photons off of hot coronal
electrons (for an excellent review of the observations, see
\cite{mccli04}). 

Motivated by these observations, we extend a
simple geodesic hot spot model \cite{schni04, schni05} to include
Monte Carlo scattering of photons emitted isotropically in the hot
spot's rest frame, which then propagate through a corona surrounding
the black hole. The hot spot has a planar orbit near the inner-most
stable circular orbit (ISCO) and the
corona is modeled with a self-similar density profile described by the
advection dominated accretion flow (ADAF) model \cite{naray94}. 

In Section \ref{classical_scattering}, we describe the physics of
classical electron scattering of unpolarized light, and show how
energy is transferred from high-energy electrons to
low-energy photons. Section \ref{rel_implementation} shows how this
scattering is treated in a relativistic context in the Kerr
metric. Assuming a thermal emitter with $T_{\rm em} = 1$ keV, we show
in Section \ref{effect_spectra} how the observed spectrum is modified
by the hot electrons. The shape of this modified thermal spectrum can in
turn be used to infer the properties of the corona. 

Section \ref{effect_lightcurves} shows the time-dependent effects of
scattering on the X-ray light curve, particularly how the amplitude of
modulation is damped for larger optical depths. Furthermore, we find
this damping is more significant for photons experiencing multiple
scattering events, which tend to have higher energies. We conclude in
Section \ref{implications_QPO}  with a summary of the implications these
results have for the hot spot model for QPOs.

\section{ELECTRON SCATTERING}\label{classical_scattering} 
Following Rybicki \& Lightman \cite{rybic79}, we use the cross section for
Thomson scattering of unpolarized radiation off of nonrelativistic
electrons:
\begin{equation}\label{cross_unpol}
\frac{d\sigma_T}{d\Omega} = \frac{r_0^2}{2}(1+\cos^2\theta),
\end{equation}
where $r_0$ is the classical electron radius $r_0 = 2.82\times
10^{-13}$ cm.

It is important to note that ``nonrelativistic'' is
a reference to the \textit{photon} energy, not the electron energy. In
the electron rest frame, we require $h\nu \ll m_ec^2$ in order for the
above cross sections to be valid, in which case the scattering is nearly
elastic or \textit{coherent}. For higher energy photons, the
scattering involves quantum effects and requires the ``Klein-Nishina''
cross section \cite{heitl54}. Since we are primarily
interested in the scattering of
photons from a relatively cool thermal accretion disk ($h\nu \sim 1-5$
keV), the classical treatment should suffice.

Even if the scattering is treated as coherent in the electron frame,
in the lab frame energy can be (and often is) transferred from the
electron to the photon. To see this boosting effect, consider a photon
with initial energy $\varepsilon_i$ scattering off an electron with velocity
$\beta$ in the $x$-direction in the ``lab frame'' $K$. In this frame,
the angle between the incoming photon and electron velocity is
$\theta$. In the electron rest frame $K'$, the photon is scattered
at an angle $\theta'$ with respect to the
$x'$-axis. The Doppler shift formula \cite{rybic79} gives
\begin{eqnarray}
\varepsilon_i' &=& \varepsilon_i \gamma(1-\beta\cos\theta) \nonumber\\
\varepsilon_f &=& \varepsilon_f' \gamma(1+\beta\cos\theta'),
\end{eqnarray}
where $\gamma = 1/\sqrt{1-\beta^2}$ and 
$\varepsilon_f$ is the post-scattering energy in the lab frame. In the
electron frame, we assume elastic scattering with
$\varepsilon_i'=\varepsilon_f'$, which should be the case for the
typical seed photons from a thermal emitter at $T_{\rm em} \sim 1$ keV.

Averaging over all angles $\theta$ (isotropic electron distribution)
and $\theta'$ [weighted by eqn.\ (\ref{cross_unpol})], we
find that the typical scattering event boosts the photon energy by
\begin{equation}\label{ef_ei}
\frac{\varepsilon_f}{\varepsilon_i} \approx \gamma^2.
\end{equation}
To determine $\gamma$, we consider a Maxwell-Boltzmann distribution
function in electron momentum $p=\gamma m v$: 
\begin{equation}\label{maxwellian}
f(p)d^3\mathbf{p} \propto 4\pi p^2 \exp\left(-\frac{\sqrt{p^2c^2 +
m_e^2c^4}}{kT_e}\right).
\end{equation} 

\section{RELATIVISTIC IMPLEMENTATION}\label{rel_implementation}
Unlike the approach taken in \cite{schni04}, where the
photons were traced backwards in time from a distant observer to the
emitting region, here it is conceptually easier to trace the photons
\textit{forward} in time from the emitter to the observer, then use
Monte Carlo methods to determine the distribution of scattered
photons. For the Thomson cross section, the path of each
photon is energy-independent,
so the photon's final observed energy can be thought of as a fiducial
redshift $E_{\rm obs}/E_{\rm em}$ that can be convolved with the
spectrum in the local emitter frame to produce the total spectrum seen
by the observer.

To determine the initial momentum of each photon, we construct a
tetrad centered on the emitter's rest frame, denoted by tilde
indices $\tilde{\mu}$. In the coordinate basis,
the energy and angular momentum of a particle on a
stable circular orbit around a Kerr black hole are given by
\begin{equation}
-p_t({\rm em}) = \frac{r^2-2Mr\pm a\sqrt{Mr}}{r(r^2-3Mr\pm 2a\sqrt{Mr})^{1/2}}
\end{equation}
and
\begin{equation}
p_\phi({\rm em}) = \pm\frac{\sqrt{Mr}(r^2\mp 2a\sqrt{Mr}+a^2)} {r(r^2-3Mr\pm
 2a\sqrt{Mr})^{1/2}}.
\end{equation}
From these we construct the 4-velocity via the inverse metric
$p^\mu({\rm em}) = g^{\mu \nu} p_\nu({\rm em})$, which gives
$\mathbf{e}_{\tilde{t}}$. Then $\mathbf{e}_{\tilde{r}}$ and
$\mathbf{e}_{\tilde{\theta}}$ are defined parallel to the coordinate
basis vectors $\mathbf{e}_r$ and $\mathbf{e}_\theta$, and
$\mathbf{e}_{\tilde{\phi}}$ is given by orthogonality. 
 
In this basis, the initial photon direction is picked randomly from an
isotropic distribution, uniform in spherical coordinates
$\cos\tilde{\theta}=[-1,1]$ and $\tilde{\phi}=[0,2\pi)$. All photons
are given the same initial energy in 
the emitter frame $p_{\tilde{t}} = -E_0$, which is used as a reference
energy for calculating the final redshift with respect to a stationary
observer at infinity. Given $p_\mu$, the photon's
geodesic trajectory is integrated using the Hamiltonian formulation
described in \cite{schni04}.

\begin{figure}[t]
\begin{center}
\scalebox{0.5}{\includegraphics*[74,330][540,750]{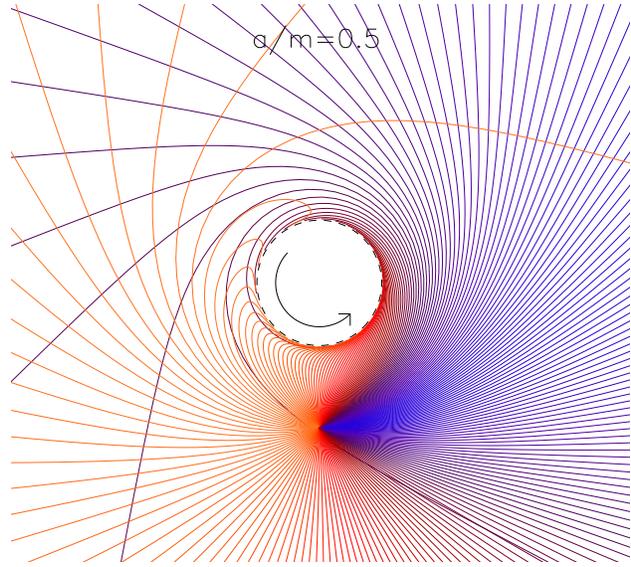}}
\caption[Photons traced from isotropic emitter at
ISCO]{\label{emitter_rays} Planar photons emitted isotropically in the
rest frame of a massive particle on a circular orbit at the ISCO.
The photon
paths are colored according to their red- or blue-shift in energy with
respect to $E_0$ measured in the emitter's frame.}
\end{center}
\end{figure}

Figure \ref{emitter_rays} shows an ``overhead view'' of photon
trajectories in the plane of the disk, emitted isotropically by a
massive test particle on a circular orbit at the ISCO of a black hole
with $a/M=0.5$. The photons are colored according to their
energy-at-infinity $E_{\rm obs}= E_\infty=-p_t$, either 
blue- or redshifted with respect to their energy in the emitter frame
$E_0$. The relativistic beaming is done
automatically by the Lorentz boost from the emitter to the coordinate
frame, so the blue photons are clearly bunched more tightly
together, as required by the invariance of $I_\nu/\nu^3$. 

At each step along the photon's path, we determine the probability of
electron scattering according to the differential optical depth
$d\tau_{\rm es} = \kappa_{\rm es}\rho ds.$
The density $\rho$ is defined in the ``Zero Angular Momentum
Observer'' (ZAMO) frame \cite{barde72} and the opacity
$\kappa_{\rm es}$ is given by the classical cross section
quoted above in equation (\ref{cross_unpol}). The scattering is then
treated classically in the electron's local frame, after which the new
photon 4-momentum is transformed back to the coordinate basis and the
ray-tracing continues along the new trajectory. 

\section{EFFECT ON SPECTRA}\label{effect_spectra}
As we showed at the end of the Section \ref{classical_scattering}, one
effect of the coronal scattering
is generally a transfer of energy from the electron to
the photon. One way to quantify this energy transfer is through the
Compton $y$ parameter, defined as the average fractional energy change
per scattering, times the number of scatterings through a finite
medium. For nonrelativistic electrons, Rybicki \& Lightman
\cite{rybic79} show that the average energy transfer per scattering
event is
\begin{equation}\label{ef_ei2}
\frac{\varepsilon_f-\varepsilon_i}{\varepsilon_i} =
\frac{4kT_e}{m_ec^2}. 
\end{equation}

The mean number of scatterings for an optically thin medium is simply
$\tau_{\rm es}$, the total optical depth through the medium. For
optically thick systems, the photons must take a random walk to
escape, so the number of scatterings becomes $\tau^2_{\rm es}$. Thus the
Compton $y$ parameter for a finite medium of nonrelativistic electrons
is 
\begin{equation}\label{compton_y}
y = \frac{4kT_e}{m_ec^2} \mbox{Max}(\tau_{\rm es},\tau_{\rm es}^2).
\end{equation}
For a low-energy soft photon source with multiple scattering events,
the final spectrum
due to inverse-Compton scattering can be estimated using the
\textit{Kompaneets equation}. For $kT_{\rm em} \lesssim h\nu \lesssim
kT_e$, the resulting spectrum takes the power-law form
\begin{equation}
I_\nu \sim \nu^{-\alpha},
\end{equation}
with
\begin{equation}\label{powerlaw_alpha}
\alpha = \frac{3}{2}+\sqrt{\frac{9}{4}+\frac{4}{y}}.
\end{equation}
At energies above $kT_e$, the electrons no longer efficiently transfer
energy to the photons, so the spectrum shows a cutoff for $h\nu \gtrsim
kT_e$:
\begin{equation}
I_\nu \sim \nu^3 \exp(-h\nu/kT_e).
\end{equation}

With the assumption of purely elastic scattering, we cannot
actually reproduce this cutoff effect; all photons are scattered
equally, and the ratio $\varepsilon_f/\varepsilon_i$ is
independent of energy. Thus equation (\ref{ef_ei2}) would predict
infinite energy boosts until $h\nu \gg m_ec^2$. In reality, higher
energy photons tend to lose
energy in scattering, due to the recoil of the electron. This effect
is relatively easy to calculate from conservation of energy and
momentum in the electron rest frame:
\begin{equation}\label{energy_compton}
\varepsilon_f = \frac{\varepsilon_i}{1+\frac{\varepsilon_i}{m_ec^2}
(1-\cos\theta)}
\end{equation}
To accurately include this effect, we would have to keep track of the
real ``physical'' energy of each photon, instead of the fiducial
redshift method that we currently use to reconstruct the total
spectrum afterwards. Ultimately, this is just a matter of
computational intensity and no real conceptual difficulty. To
first-order, we can treat the thermal photon source as a
monochromatic emitter at $E_0 = 3kT_{\rm em}$, which gives a
reasonable approximation to the true solution.

For the corona geometry, we use a self-similar distribution described
by an ADAF model
\cite{naray94}, with density and temperature profiles that scale as
\begin{eqnarray}
\rho &\propto& r^{-3/2}, \\
T &\propto& r^{-1}
\end{eqnarray}
outside of the ISCO. We have ignored the bulk velocity of the inwardly
flowing gas, which will typically have $v_{\rm bulk} \ll v_{\rm
therm}$ in the ADAF model.

Without scattering, the
time-averaged ``numerical'' spectrum could be described by the
relativistic transfer function described in \cite{schni04}, defined
over an infinitesimal band in radius
$R_{\rm in} \approx R_{\rm out}=r_{\rm em}$. The
inverse-Compton processes in the corona serve to further broaden this
transfer function, as in \cite{sunya80} and
\cite{titar94}. This transfer function is then normalized to
the rest energy $E_0$ and convolved with the actual emission spectrum
(e.g.\ a thermal blackbody at $kT_{\rm em}$) to give the simulated
observed spectrum. 

\begin{figure}[t]
\begin{center}
\scalebox{0.5}{\includegraphics*[80,360][540,700]{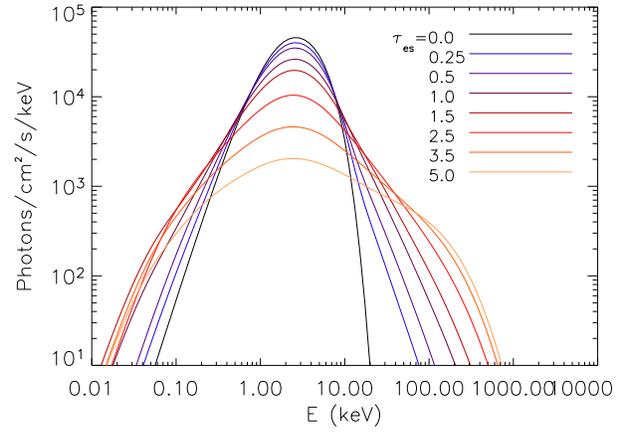}}
\caption[Spectra of a thermal source scattering through a hot
corona]{\label{powerlaw_cutoff} Simulated observed spectra of a
thermal hot spot emitter with $T_{\rm em}=1$ keV, on a circular orbit
at the ISCO of a black hole with $a/M=0.5$. The thermal spectrum is
modified by relativistic effects and Compton scattering off a hot
corona with $T_e = (r_{\rm ISCO}/r)100$ keV.}
\end{center}
\end{figure}

Figure \ref{powerlaw_cutoff} shows a set of these simulated spectra
from a hot spot emitter around a black hole with $a/M=0.5$. The
emission spectrum is thermal in the hot 
spot rest frame with $T_{\rm em}=1$ keV. The coronal ADAF model has
$T_e = (r_{\rm ISCO}/r)100$ keV, and electron density $n_e \sim
r^{-3/2}$ for a variety of optical depths $\tau_{\rm es}$. The spectra
are plotted in units of (\#photons/s/cm$^2$/keV), as is the convention
by many observers, but the actual magnitude of the $y$-axis is
arbitrary, and would normally depend on the distance to the
source.
From the slope of the power-law and the location of the cut-off, the
corona temperature and optical depth can be inferred from
observations \cite{pozdn77,sunya79,gilfa94,gierl97}. 

\section{EFFECT ON LIGHT CURVES}\label{effect_lightcurves}
\begin{figure*}
\centering
\scalebox{0.3}{\includegraphics*[-24,300][415,720]{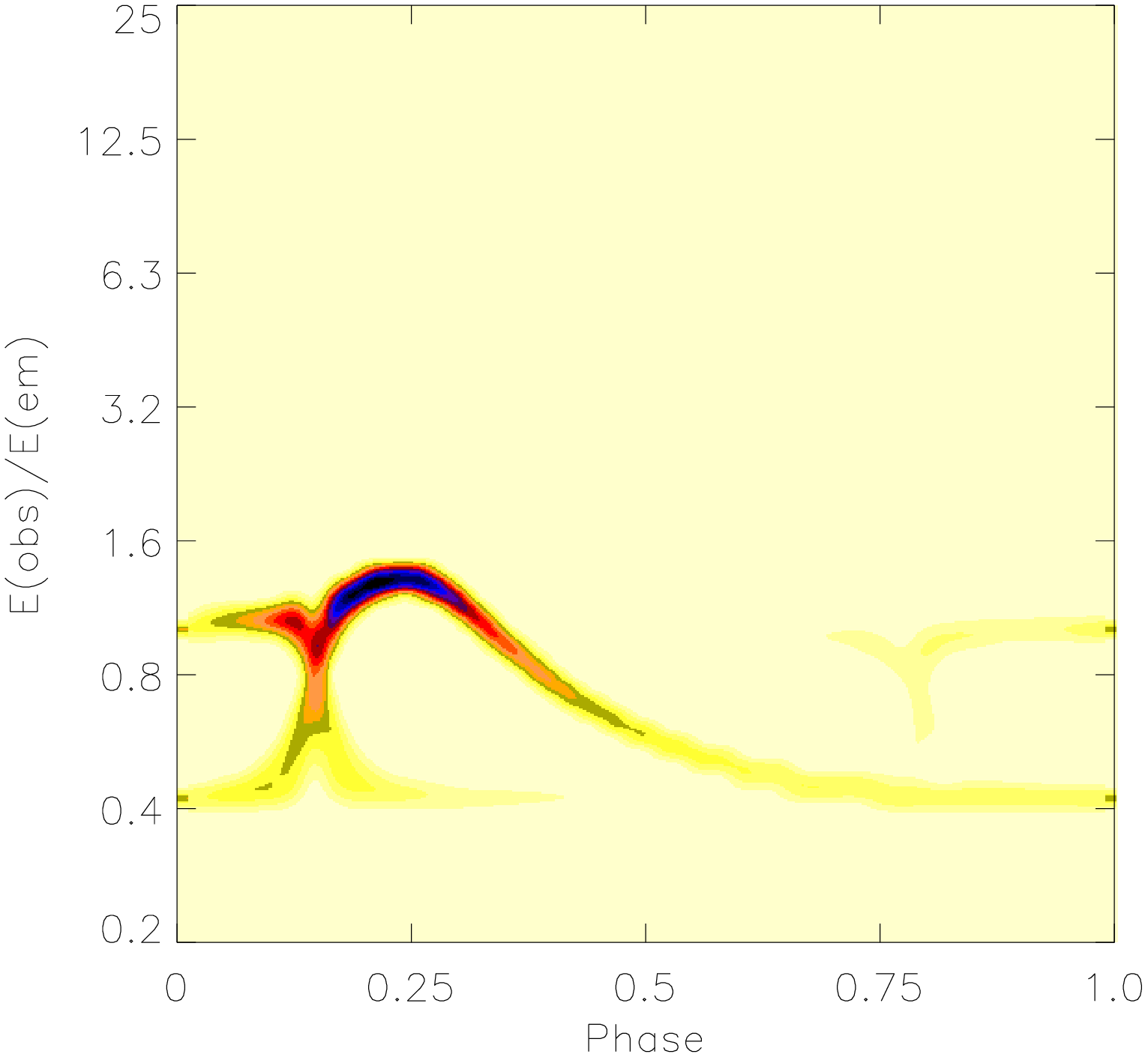}}
\scalebox{0.3}{\includegraphics*[50,300][415,720]{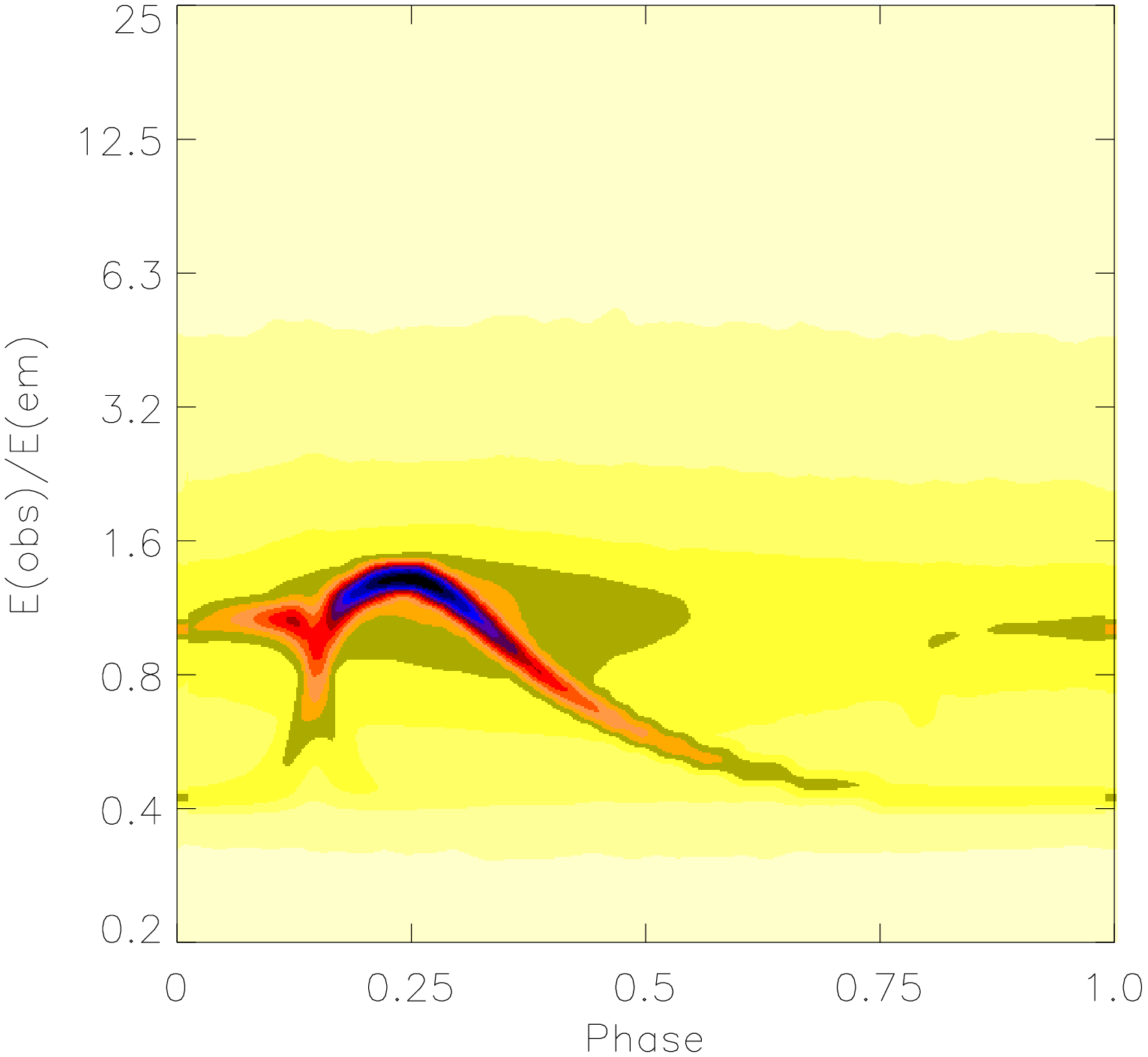}}
\scalebox{0.3}{\includegraphics*[50,300][415,720]{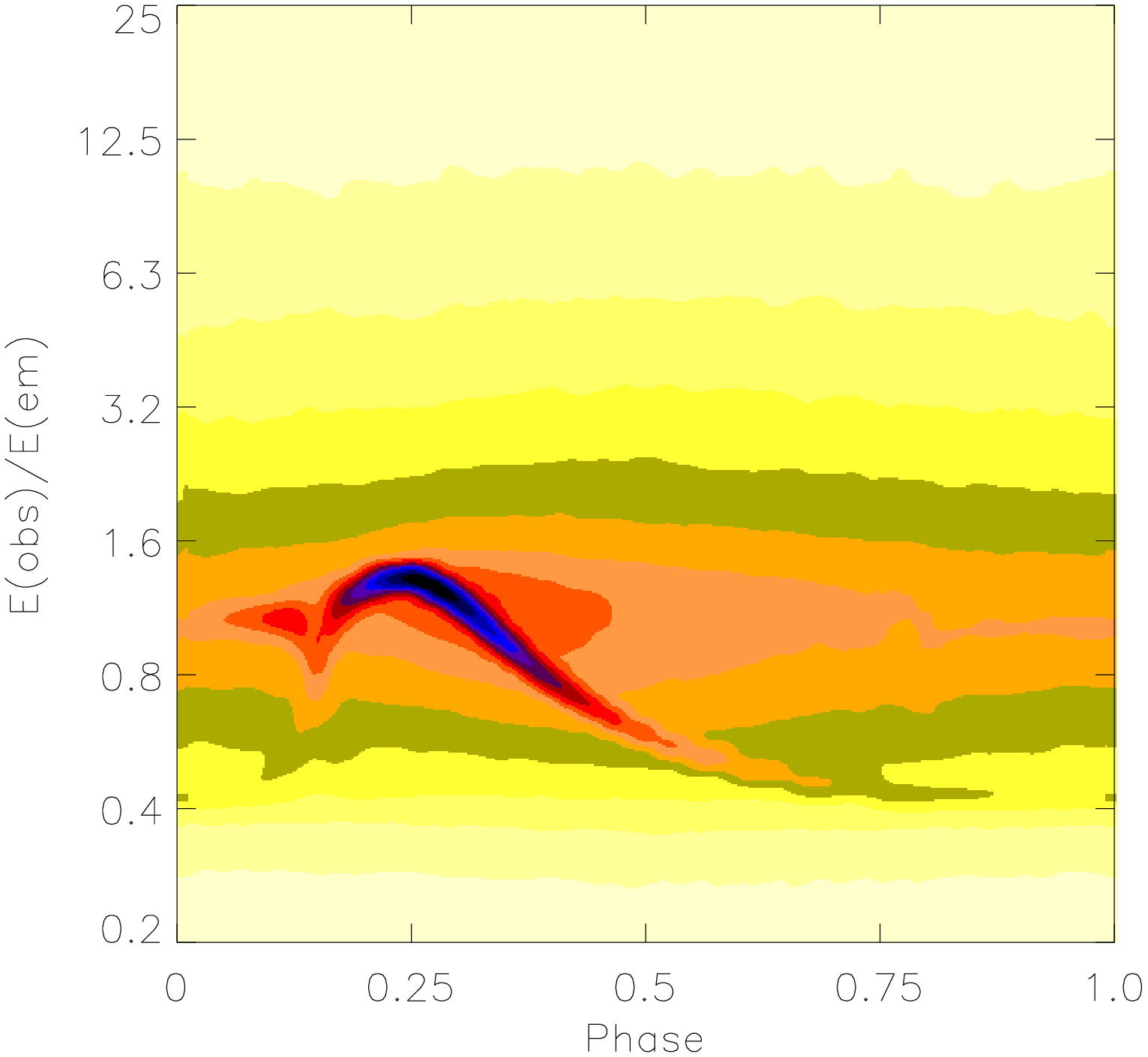}}
\scalebox{0.3}{\includegraphics*[50,300][415,720]{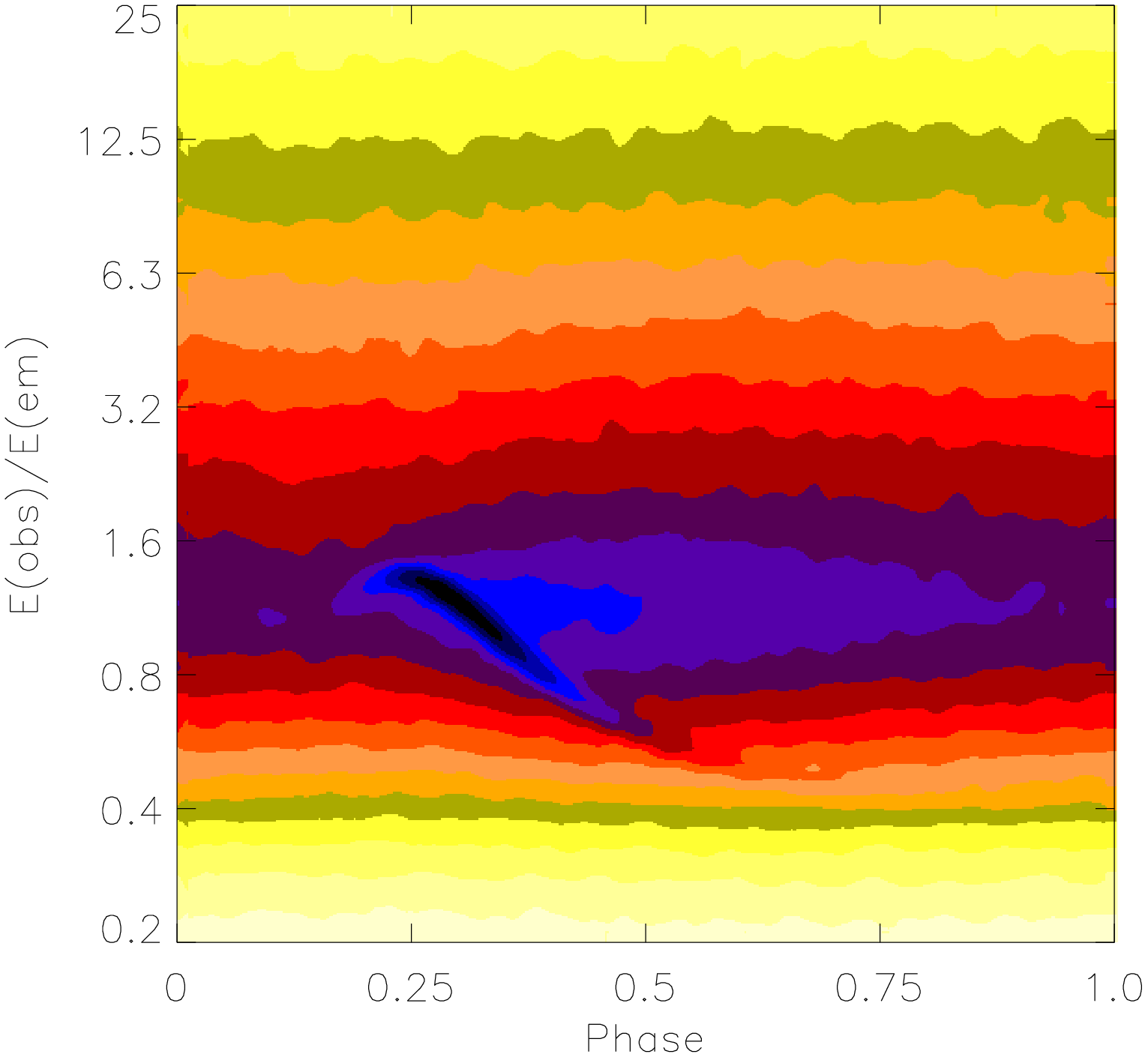}}
\caption[Spectrograms of hot spot emission including
scattering, $i=45^\circ$]{\label{scatter_spectrogram75} Time-dependent
spectra of a monochromatic, isotropic hot spot emitter on an ISCO
orbit with $a/M=0.5$ and inclination angle $i=75^\circ$. The four
panels show spectrograms for systems of
increasing optical depth $\tau_{\rm es}=[0,1,2,4]$. The logarithmic
color scale shows the number
of photons in each time/energy bin, normalized to the peak value for
each panel.}
\end{figure*}

The spectra in Figure \ref{powerlaw_cutoff} were created by
integrating over the complete hot spot orbital period and over all
observer inclination angles. However, during the Monte Carlo
calculation, it is just as simple to bin all the photons according to
their final values of $\theta$, $t_{\rm obs}$, and energy $-p_t$.
The latitude bins are evenly spaced in
$\cos\theta$ so that a comparable number of photons land in each
zone. The energy bins are spaced logarithmically to include the high
energy tail and also maintain high enough spectral resolution at lower
energies. Assuming the hot spot is on a circular periodic orbit, the
photons detected at any azimuthal position can be mapped into the
appropriate bin in $t_{\rm obs}$, so that none are ``wasted.''

An excellent way to see the effects of scattering on the hot spot
light curves is by plotting time-dependent spectra of a monochromatic
emitter, shown in Figure \ref{scatter_spectrogram75} for a range of
optical depths and an inclination angle of $75^\circ$. 
The logarithmic color scale shows (\#photons/s/cm$^2$/keV/period),
normalized to the peak intensity in each panel. 
At ``0'' phase, when the emitter is on the far side of the black hole,
the spectrum shows two distinct lines, one
blueshifted in the forward direction of hot spot motion, and one
redshifted in the backward direction. As the hot spot comes around
towards the observer, the directly beamed blueshifted line dominates,
and then when the phase is $\sim 0.5$ and the hot spot is on the near
side of the black hole, a single line dominates. This is due to the
gravitational demagnification of the secondary images formed by
photons that have to complete a full circle around the black hole to
reach the observer. While these features would most likely be
unresolvable for black hole binaries, they may well be observable in
X-ray flares from Sgr A$^\ast$ as well as other supermassive black
holes (e.g.\ see \cite{bagan01}).

In the subsequent panels, the spectrum is modified by the scattering
of the hot spot photons in the surrounding corona. As in Section
\ref{effect_spectra}, the temperature and density profile of the
corona is given by an ADAF model with $T_e(r_{\rm ISCO})=100$ keV. The
four panels of Figure \ref{scatter_spectrogram75} show increasing
values of $\tau_{\rm es}=[0,1,2,4]$. The effects of scattering on the
spectra are really quite profound. As we described qualitatively in
\cite{schni05}, the electron corona is like a cloud of fog surrounding a
lighthouse, spreading out the delta-function beam in time and
energy. 

Unlike the simple model there, where each photon was assigned some
positive time delay, the full Monte Carlo scattering calculation 
shows that some photons actually arrive \textit{earlier} in time by
taking a ``shortcut'' to the observer instead of waiting for the hot
spot to come around and move towards the observer. And of course, the
photons are also spread out in energy due to the inverse-Compton
effects. As the optical depth increases, the well-defined curves in
Figure \ref{scatter_spectrogram75}a is smeared out into a nearly
constant blur when $\tau_{\rm es} =4$, with a broad spectral peak as in
Figure \ref{powerlaw_cutoff}. Only a slight trace of the original
coherent light curve remains, composed of roughly $1\%$ of the emitted
photons that do not scatter before reaching the observer or get
captured by the event horizon. When $\tau_{\rm es} > 1$, multiple
scattering become more common, so photon shortcuts become rarer,
tending to spread the light curve preferentially to the right (delay in
observer time), as seen in Figures \ref{scatter_spectrogram75}c,d.

\begin{figure}[ht]
\begin{center}
\includegraphics[width=0.5\textwidth]{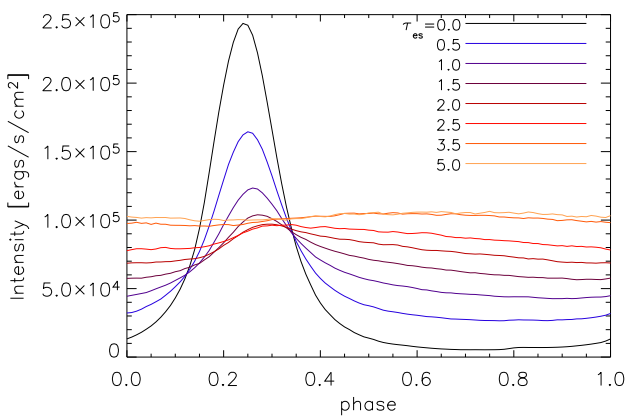}
\caption[Integrated light curves for a range of optical
depths]{\label{lightcurves_tes} Energy-integrated light curves for a
hot spot with orbital parameters as in Figure
\ref{scatter_spectrogram75} and a range of coronal optical depths. The
emitted spectrum is assumed to be
thermal with a hot spot temperature $T_{\rm hs}=1$ keV, integrated
over $0.5-30$ keV in the observer's frame.}
\end{center}
\end{figure}

By integrating over broad energy bands such as those typically used in
\textit{RXTE} observations, we can increase our ``signal'' strength
while sacrificing spectral resolution. For millisecond periods, there
will still not be nearly enough photons to provide phase resolution,
but these features may show up statistically in the power spectrum or
bispectrum \cite{macca04}. Figure \ref{lightcurves_tes}
shows a set of integrated light curves for a variety of optical
depths. The black hole and hot spot parameters are as in Figure
\ref{scatter_spectrogram75}, here assuming a thermal hot spot emitter
with temperature $T_{\rm hs}=1$ keV. As the optical depth to electron
scattering increases, the rms 
amplitude of each light curve decreases as the photons get smoothed
out in time. Similarly, due to the average time delay added to each
photon by the increased path length, the relative location of each
peak is shifted later in time. 

In all likelihood, the relative phase shifts would be nearly
impossible to detect, regardless of the instrument sensitivity, since
to do so would require measuring the light curve from a single
coherent hot spot at two different optical depths. It is difficult to
imagine a scenario where the coronal properties could change on such
short time scales (yet it is possible that a fixed hot spot on the
surface of an X-ray pulsar might actually be used for this
technique). However, the higher harmonic peaks of the different
light curves may in fact be measurable with the next-generation X-ray
timing mission, or under extremely favorable conditions, even with
\textit{RXTE}. 

\begin{figure}[ht]
\begin{center}
\includegraphics[width=0.45\textwidth]{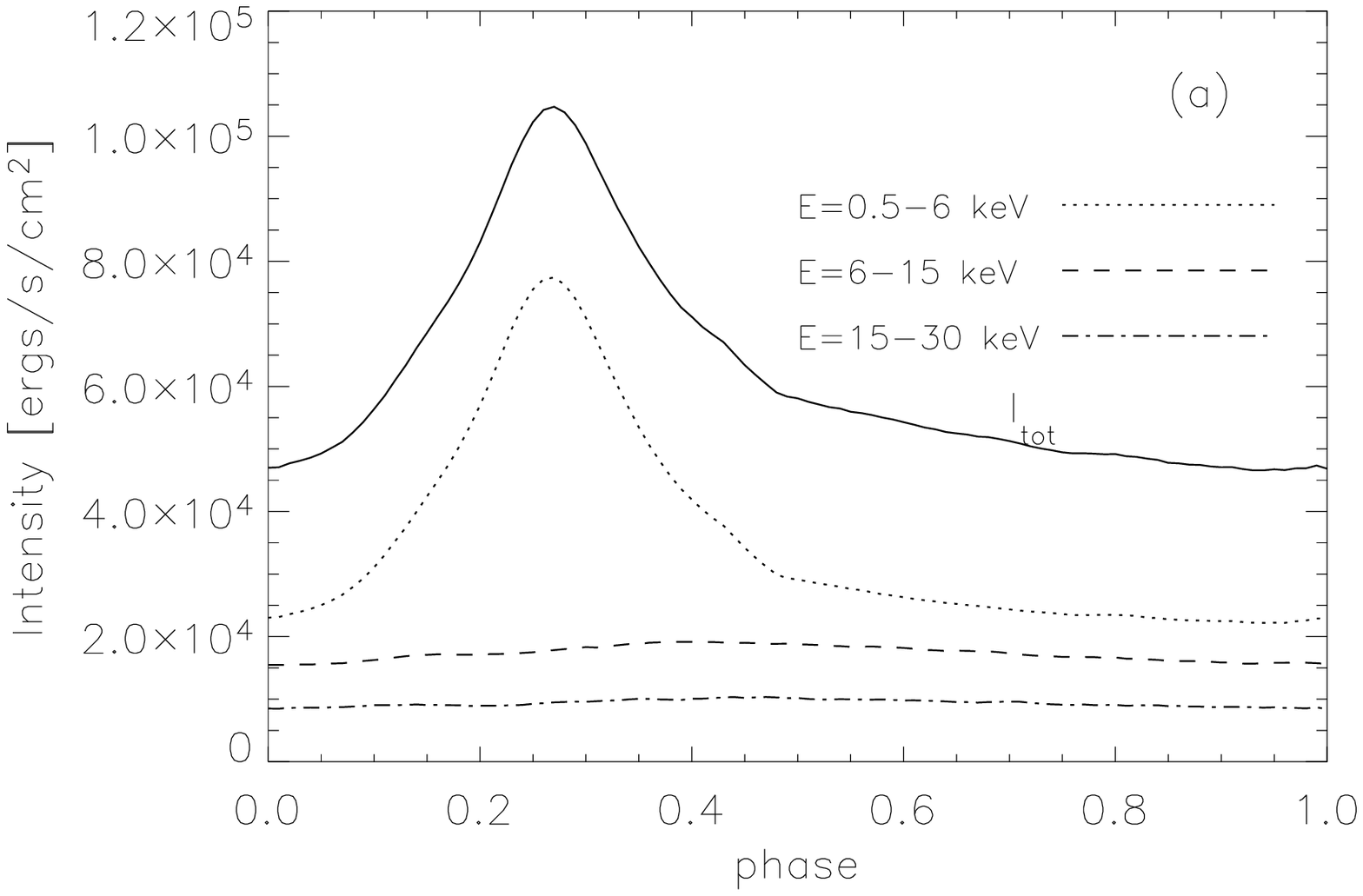}
\includegraphics[width=0.45\textwidth]{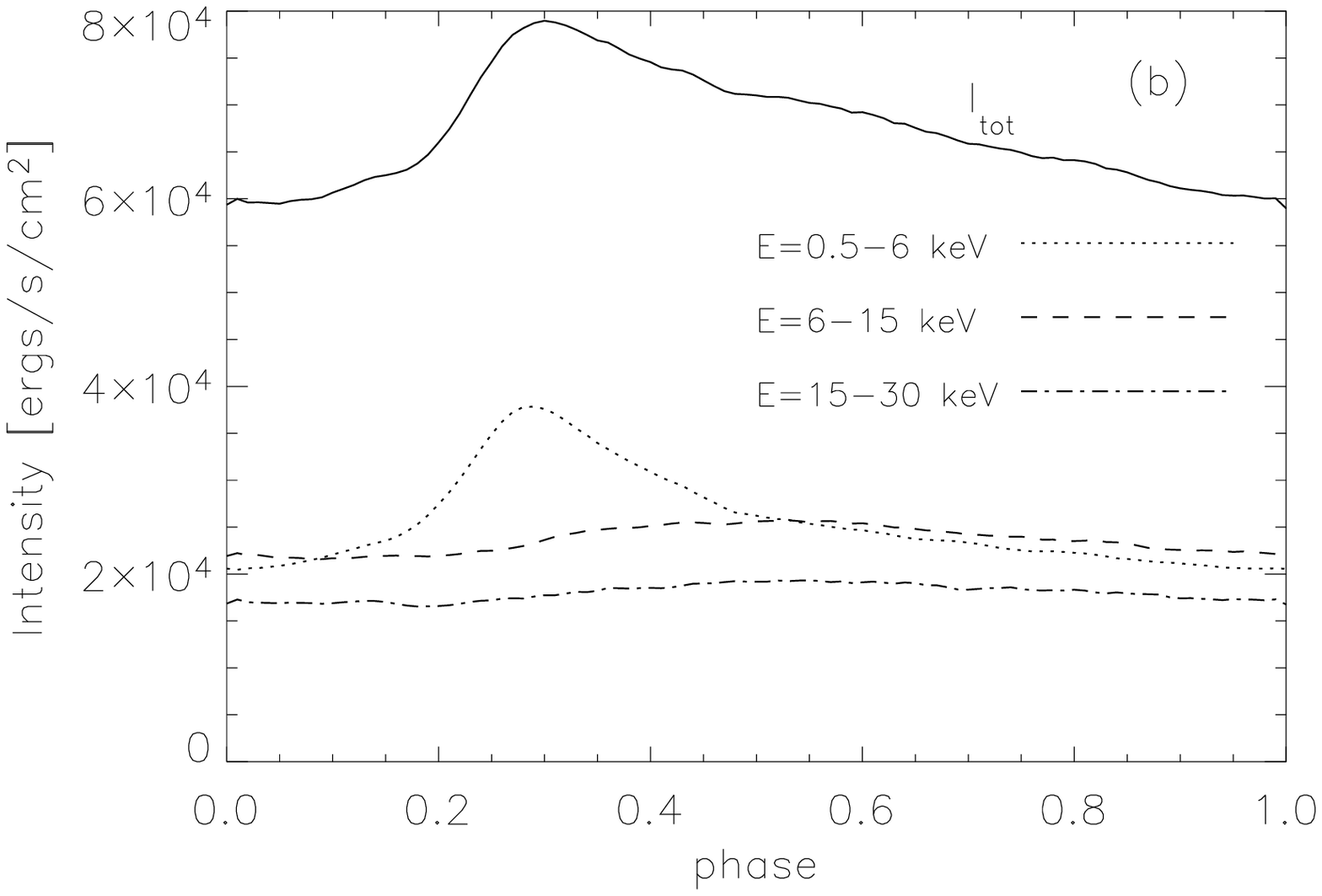}
\caption[Light curves in different \textit{RXTE} energy
bands]{\label{lightcurves_eband} Hot spot light curves in a few
different \textit{RXTE} energy bands (we have expanded the lowest
energy band down to 0.5 keV to include the thermal emission of a hot
spot at $T_{\rm hs}=1$ keV). The hot spot inclination is $75^\circ$
and the coronal properties are as in Figure
\ref{scatter_spectrogram75}. The optical depth to scattering is
$\tau_{\rm es}=1.5$ in (a) and $2.5$ in (b). The higher energy light
curves are made from photons that have experienced more scattering
events, boosting their energy and delaying their arrival time.}
\end{center}
\end{figure}

While the absolute peak shifts for hot spot light curves at different
optical depths would probably not be detectable, the relative shifts of
simultaneous light curves in different energy bands may be observable,
at least on a statistical level with a cross-correlation
analysis. Since the average scattering event boosts photons to higher
energy bands and also causes a net time delay due to the added
geometric path, the light curves in higher energy bands should be
delayed with respect to the lower energy light curves. A few of the
typical energy bands used for \textit{RXTE} observations are $(2-6)$,
$(6-15)$, and $(15-30)$ keV. To fully cover the peak emission from a
thermal hot spot at 1 keV, we expand the lowest energy band in our
calculations to cover $(0.5-6)$ keV. The light curves in these three
bands are plotted in Figure \ref{lightcurves_eband} for
$i=75^\circ$. The low energy band resembles the unscattered light
curve plus a roughly flat background, while the higher energy light
curves show a much smaller modulation with a significant phase shift
($\sim 0.3$ periods) due to the additional photon path lengths. 

\section{IMPLICATIONS FOR QPO MODELS}\label{implications_QPO} 

The original motivation for the application of scattering to the hot
spot model was to answer a few important questions raised by
\textit{RXTE} observations:
\begin{itemize}
\item The distinct lack of power in higher harmonics at
integer multiples of the peak frequencies.
\item The larger significance of high frequency QPO detections in the
higher energy bands (6-30 keV) relative to the signal in the lower
energy band (2-6 keV).
\item The trend for these HFQPOs to exist
predominantly in the Steep Power Law (SPL) spectral state of the
black hole.
\end{itemize}

Beginning with the final point, it appears to be quite reasonable that
the physical mechanism producing the power law region of the spectrum is the
inverse-Compton scattering of cool, thermal photons off of hot coronal
electrons. The \textit{steep} power law suggests a small-to-moderate
value for the Compton $y$ parameter, inferred from equation
(\ref{powerlaw_alpha}), in the range $0.5 \lesssim y \lesssim 10$. From
equation (\ref{compton_y}), this suggests either a small optical depth
or a small electron temperature. To gain insight into which of these
two options is more likely, we need to address the other two
observational clues.

In \cite{schni05}, we first proposed the scattering model as an explanation
for harmonic damping. With the more
careful treatment in this paper, we include not only
the temporal, but also the spatial effects of electron scattering. The
photons originally beamed toward the observer are now scattered in the
opposite direction, while the photons emitted away from the observer
can now be scattered back to him. This smoothes out the light curve in
time more effectively than the localized convolution functions used in
\cite{schni05}. At the same time, the scattering is not completely
isotropic [see eqn.\ (\ref{cross_unpol})], so some modulation
remains. Thus, to maintain a significant modulation in the observed light 
curve, we require a relatively small optical depth, reducing the
smoothing effects of the scattering.

The fact that most HFQPOs appear more significantly in higher energy
bands also points towards Compton scattering off hot
electrons. However, as the calculations above show (see Fig.\
\ref{lightcurves_eband}), with the basic thermal disk/hot spot model,
the light curves actually have \textit{smaller} amplitude fluctuations
in the higher
energy bands, as these scattered photons get smoothed out more in
time. Furthermore, while the higher harmonic modes are successfully
damped in the scattering geometry, so is the fundamental peak. Thus,
in order to agree with observations, the hot spot overbrightness would
need to be much higher than the values quoted in \cite{schni04,schni05}. 

Based on these arguments alone, we find it unlikely that the
HFQPOs are coming from a cool, thermal hot spot getting upscattered by
a hot corona. From the photon continuum spectra of the SPL state,
there appears to be a hot corona with $T_e \sim 100$ keV, but as
Figure \ref{lightcurves_eband}a shows, the lowest energy photons,
which presumably do not scatter in the corona,
have by far the greatest amplitude modulations. It is possible that the
relative modulation would appear smaller due to the added flux from
the rest of the cool, thermal disk, but much of this steady-state
emission should
also get scattered to higher energies, further damping the modulations
in the $(6-30)$ keV bands. 

The high luminosity of the SPL state (also called the Very
High state) suggests that the thermal, slim disk geometry may not be
appropriate here. Perhaps it is more likely that these cases
correspond to an ADAF model, traditionally associated with very low or
very high accretion rates \cite{naray94}. Since the ADAF model cannot
radiate energy
efficiently, the gas in the innermost regions will be much hotter than
in the slim disk paradigm. Thus \textit{hot} hot spots with $T_{\rm hs}
\gtrsim 5$ keV could be forming inside a
small ADAF coronal region, providing seed photons that are already in
the higher energy bands, and are only moderately upscattered by the
surrounding corona.

Future work will focus on exploring other corona geometries and
applying the Monte Carlo scattering code to other QPO models in order
to better understand the emission processes that describe the SPL
spectral state.

\bigskip 
\begin{acknowledgments}
We would like to thank Ed Bertschinger and Ron Remillard for many
helpful discussions. This work was supported by NASA grant
NAG5-13306.
\end{acknowledgments}

\bigskip 

\end{document}